# CareerMapper: An Automated Resume Evaluation Tool

Vivian Lai, Kyong Jin Shim, Richard J. Oentaryo, Philips K. Prasetyo, Casey Vu
Ee-Peng Lim, David Lo
School of Information Systems, Singapore Management University, Singapore
{vivianlai, kjshim, roentaryo, pprasetyo, caseyanhthu, eplim, davidlo}@smu.edu.sg

*Abstract*—The advent of the Web brought about major changes in the way people search for jobs and companies look for suitable candidates. As more employers and recruitment firms turn to the Web for job candidate search, an increasing number of people turn to the Web for uploading and creating their online resumes. Resumes are often the first source of information about candidates and also the first item of evaluation in candidate selection. Thus, it is imperative that resumes are complete, free of errors and well-organized. We present an automated resume evaluation tool called "CareerMapper". Our tool is designed to conduct a thorough review of a user's LinkedIn profile and provide best recommendations for improved online resumes by analyzing a large number of online user profiles.

## I. INTRODUCTION

Finding a job is one of the key tasks that people perform today. Almost all companies today leverage online platforms and services for harvesting and managing candidates' information. A number of online resume distribution and search sites have widely been in use worldwide today, such as CareerBuilder, Indeed, Glassdoor, Monster, LinkedIn, etc. LinkedIn is undoubtedly one of the most popular social networking platforms in the world for business networking with over 300 million members [1]. Despite its fast growth, however, only half of its members are known to have "complete" profiles [2]. Half of the users are not getting the results they would like (such as calls from recruiters about better job opportunities) because they have not spent enough time to understand and state clearly what should be on their profile. Having a complete, error-free and well-organized resume is a key to leaving a good first impression with recruiters and employers [3]. Hence, it is important to check one's resume carefully.

## II. AUTOMATED RESUME EVALUATION

For busy students and professionals, often the important task of creating, updating and maintaining a resume can come across as a daunting task. Traditionally, for paper-based resumes, candidates would visit resume clinics or resume experts for help. Upon completing the first draft, the candidate would sit down with a resume expert and have each section reviewed for correctness and so on. One can do this in a more crowdsourcing-based manner whereby a number of acquaintances with sufficient professional experience are asked to review a candidate's resume and provide feedback. Another approach to creating a successful resume is to model after other successful resumes. The candidate would carefully go through other model resumes and try to replicate their organization, structure and wording. These are all great ways of improving one's resume.

We propose to automate this process of resume evaluation in a much larger scale. Given a large number of up-to-date LinkedIn profiles, "CareerMapper" provides a thorough evaluation of a user's LinkedIn profile. The evaluation is performed by scanning through over 1.6 million profiles of other LinkedIn users, section by section. This is similar to how a user would check on other profiles manually, section by section. CareerMapper automates this process and performs the task programmatically in a much more time efficient manner. Recently, CareerMapper expanded its service to include another resume data source (a proprietary resume creation and search platform) and offers the same resume evaluation service for all profiles on that platform.

## III. CAREERMAPPER

### A. System Architecture

CareerMapper is a web-based application written in PHP, HTML, JavaScript and CSS. The web front serves as the main interface for the user. Once a name search is initiated by the user in a web browser, CareerMapper issues a database search query against the back-end Elasticsearch data source. This data source harvests and maintains a large data repository of LinkedIn user profiles. Elasticsearch returns the results in JSON format and CareerMapper then parses and displays the results in the web interface. CareerMapper web application runs on a Tomcat web server, and it can be accessed at: http://research.larc.smu.edu.sg/careermapper.

### B. Data Harvesting

In this work, we study the 1.6 million user profiles extracted from the LinkedIn of Singapore, all of which are stored in our Elasticsearch indices. The data consist of (1) public user profiles that can be accessed from Singapore users directory and (2) company profiles that are referred to in the public user profiles. It is worth noting that, while our OPN data cover a comprehensive set of user profiles, it may not necessarily be able to capture all sorts of occupation in Singapore. The data may leave out, for instance, blue-collar/non-technical workers who do not use social media. Nevertheless, our data are arguably representative of all professionals, managers, executives and technicians (PMET), who increasingly make up the majority of working population in Singapore [4].

For data collection, we devised a data crawler that performs three steps to extract public user profiles and company profiles from LinkedIn. Firstly, we collect all user profile URLs from the Singapore member directory. For the subsequent months, we crawl the content of each user profile using the previously collected profile URLs. Note, however, that our crawler captures neither users' connection list nor wall posts. Finally, while crawling the user profiles, the company URLs found in each user profile are collected and used to crawl the company profiles (pages).

## C. Searching for a Profile

The primary purpose of using CareerMapper is to have one's resume (or profile) professionally evaluated. This journey starts with the user searching for his own profile. In the main page of CareerMapper, if the user already has a LinkedIn profile, he can simply key in his first name and last name. The current version of CareerMapper allows the user to specify his last graduated university or educational institution. If the user's name is common and it is expected that there may be many profiles with the same name, selecting this "last graduated institution" option will narrow down the search. If the user does not currently have a live LinkedIn profile, he can simply go to the next tab "Create your account or log in now". There, the user can create a new LinkedIn profile or he can create a new profile within Living Analytics Research Centre's proprietary resume creation and search platform.

Upon issuing the name search, CareerMapper finds matching LinkedIn profiles from its Elasticsearch data source. Figure 1 shows the search results with two matching profiles. The match shown on the left hand side (with the blue LinkedIn logo) is the user's LinkedIn profile. The match shown on the right hand side (with the white colored icon) is the user's profile in our proprietary resume platform.

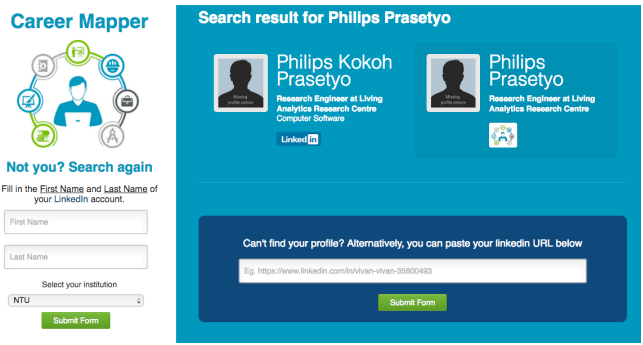

Figure 1. Matching profiles.

## D. Evaluation of Resume (Profile) & Recommendations

CareerMapper first checks on the "completeness" of the user's profile. Upon clicking on the first matched profile, the user can see the summary of suggestions at the top of the screen as shown in Figure 2. Depending on the data source from which the user's profile comes from, the specifics of what CareerMapper checks on for evaluation may vary. In the case of LinkedIn profiles, there are several sections such as the user's "Basic Information" (which includes the user's name), "Experience" (which includes one or more present and past job experiences), "Education" (which includes present and past education experiences), and so on. Each section further comprises one or more instances, e.g., if one has obtained a Bachelor's degree and a Master's degree, the "Education" section of his profile would list two Education "instances".

Figure 2 shows that CareerMapper is suggesting that the user fills out the "Award" section of the user's LinkedIn profile. The idea behind making this suggestion is that a certain percentage (this threshold value can be configured in the CareerMapper program) of other LinkedIn profiles with similar specifications as the user's indicated one or more awards in their "Award" section. The example shown in Figure 2 uses "Education" as the "similarity" criterion and "20%" as the threshold value.

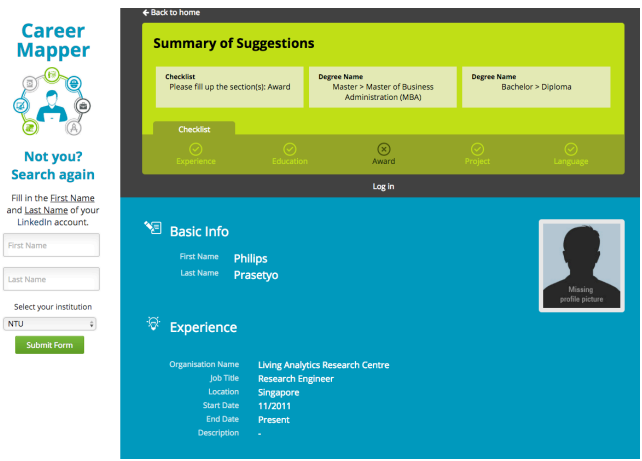

Figure 2. Summary of suggestions.

CareerMapper makes recommendations in details as shown in Figure 3. For a user who indicated "Master" as the degree name for a particular Education instance. CareerMapper does not indicate whether the user's original naming or entry is correct or incorrect. Rather, it simply provides to the user "recommendations" based on how other users indicated the similar in their Education section.

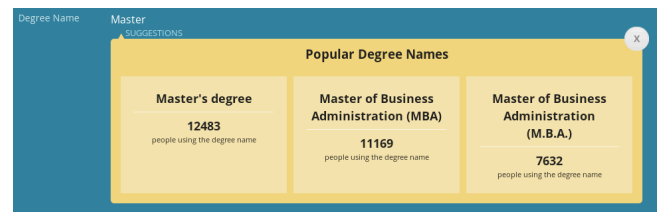

Figure 3. Suggestion for degree name.

For example, Figure 3 shows that many other users tend to use more specific names for their master's degree. Over 12,000 users indicate "Master's degree" instead of "Master". Likewise, over 11,000 users indicate "Master of Business Administration (MBA)".

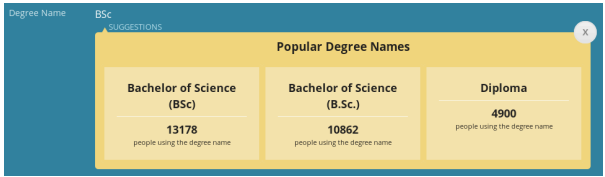

Figure 4. Suggestion for degree name.

Figure 4 shows another example of name recommendation. In this case, instead of "Bsc", CareerMapper suggests more detailed degree names such as "Bachelor of Science (BSc)" and "Bachelor of Science (B.Sc.)" because a large number of other users state longer degree names.

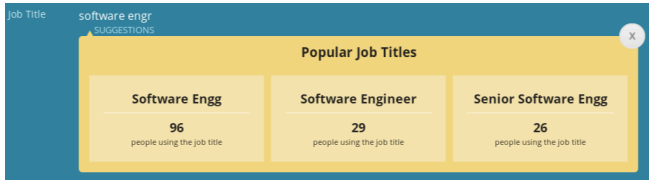

Figure 5. Suggestion for job title.

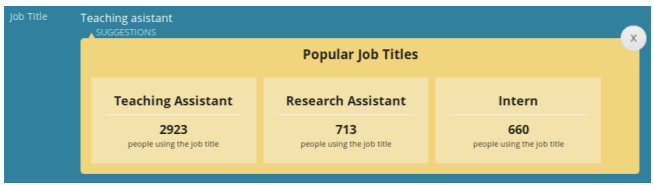

Figure 6. Suggestion for job title.

Our analysis over a large number of profiles indicates that some users do not fully spell out their job titles or (study) majors. Figure 5 shows a user specifying "software engr" as the job title for an Experience instance. This job title appears unprofessional due to lack of appropriate usage of uppercase and lowercase letters. CareerMapper finds that more appropriate titles are "Software Engg" or "Software Engineer". Additionally, as shown in Figure 6, CareerMapper performs spell checking. In the example, "Teaching asistant" is missing an 's' in the second word.

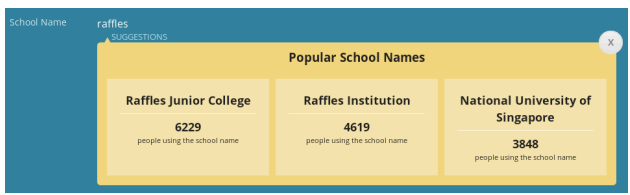

Figure 7. Suggestion for school name.

Figure 7 shows an example of CareerMapper making recommendations for school (educational institution) name. The user specified "raffles" as the institution, but this can be confusing because there are two different "Raffles" schools in Singapore: "Raffles Junior College" and "Raffles Institution". Making this entry more specific would help this user's resume appear more complete and professional.

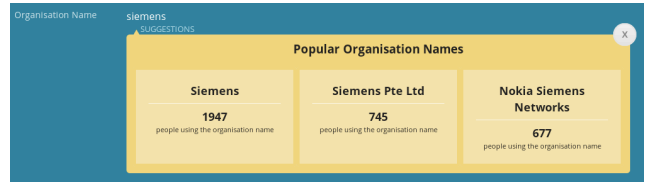

Figure 8. Suggestion for organization name.

Figure 8 shows an example of CareerMapper making recommendations for organization name (in LinkedIn's "Experience" section). CareerMapper finds top three recommendations based on usage. However, the usage (e.g. total number of profiles using each recommended name) must exceed a certain threshold and this threshold value can be configured in CareerMapper. This is to ensure that CareerMapper makes recommendations with a reasonable amount of support. In Figure 8, the lowercase letter "s" at the start of the company name can make the resume look unprofessional. Therefore, the first recommended name shows the correct company name "Siemens".

IV. SUMMARY AND FUTURE DIRECTIONS

We present an automated resume evaluation tool called "CareerMapper" and illustrate how CareerMapper evaluates professional resumes. Further, we demonstrate key examples of how CareerMapper makes recommendations for different sections of the user's resume by scanning through and deriving insights from a large pool of other resumes. Using this automated resume evaluation tool, users can quickly have their resumes evaluated and appropriate recommendations displayed in a user-friendly web interface. In the future, we plan to extend the CareerMapper framework to work with even more data sources. We also plan to develop a new career path optimization and recommendation method into CareerMapper.

ACKNOWLEDGMENT

This research is supported by the National Research Foundation, Prime Minister's Office, Singapore under its International Research Centres in Singapore Funding Initiative.